# Tuning Perpendicular Anisotropy Gradient in Co/Pd Multilayers by Ion Irradiation

Peter K. Greene,[1] Julia Osten,[2,3] Kilian Lenz,[2] Jürgen Fassbender,[2,3] Catherine Jenkins,[4] Elke Arenholz,[4] Tamio Endo,[5] Nobuyuki Iwata,[6] and Kai Liu[1,*]

[1]*Physics Department, University of California, Davis, California, 95616, USA*

[2] *Helmholtz-Zentrum Dresden-Rossendorf, Institute of Ion Beam Physics and Materials Research, , Bautzner Landstr. 400, 01328 Dresden, Germany*

[3]*Technical University Dresden, Helmholtzstr. 10, 01609 Dresden, Germany*

[4]*Advanced Light Source, Lawrence Berkeley Laboratory, Berkeley, California, 94720 USA*

[5]*Mie University, Graduate School of Engineering, Tsu, Mie 514-8507, Japan*

[6]*Department of Electronics & Computer Science, Nihon University, Chiba, 274-8501, Japan*

Abstract

The tunability of $Ar^+$ ion irradiation of Co/Pd multilayers has been employed to create depth-dependent perpendicular anisotropy gradients. By adjusting the $Ar^+$ kinetic energy and fluence, the depth and lateral density of the local structural modification are controlled. First-order reversal curve analysis through X-ray magnetic circular dichroism and conventional magnetometry studies show that the local structural damage weakens the perpendicular anisotropy near the surface, leading to a magnetization tilting towards the in-plane direction. The ion irradiation method is complementary to, and may be used in conjunction with, other synthesis approaches to maximize the anisotropy gradient.

Magnetic nanostructures with graded anisotropy offer a solution to both thermal stability and writability challenges in advanced magnetic recording media.[1-4] The interlayer exchange coupling lowers the overall coercivity, facilitating the writing process, whereas the magnetically hard layer provides pinning for the media and ensures its thermal stability. Typically the anisotropy gradient has been introduced during synthesis, e.g., by varying synthesis conditions[5-7] or individual layer thicknesses,[8] introducing composition gradients,[9-11] or inducing structural variations.[12, 13] To fully realize the benefits of graded media, it is important to maximize the anisotropy gradient over the rather limited media thickness (on the order of 10-15 nm).

Ion irradiation has been shown as an effective method in locally inducing structural imperfections/intermixing,[14-16] particularly well-suited to manipulate magnetic properties of perpendicular magnetic anisotropy systems such as Co/Pd and Co/Pt where the interface anisotropy is critical.[17, 18] The irradiation energy and fluence can be separately adjusted to control the depth and lateral extent of the damaged region, which leads to reductions in both the magneto-crystalline and interface anisotropy, and in turn modifies the sample magnetic characteristics.[19-21] These tunable characteristics are essential to precise control of the depth and strength of the anisotropy grading, providing a method to confine the gradient to a thin layer as required by modern media applications. In this work we report an alternative approach of using post-deposition $Ar^+$ ion irradiation to create an anisotropy gradient in Co/Pd multilayer films. This method is complementary to, and may be used in conjunction with, the aforementioned synthesis approaches to maximize the anisotropy gradient.

Thin films of Si substrate/Pd(20 nm)/[Co(0.3 nm)/Pd(0.9 nm)]$_{52}$/Pd(0.9 nm), with a total magnetically active thickness of 62.4 nm, were magnetron sputtered in an ultrahigh vacuum chamber with a base vacuum of $9\times10^{-9}$ torr using 5 mtorr of Ar. The as-prepared multilayers exhibit a uniform perpendicular anisotropy due to the Co/Pd interfaces. Multiple pieces of the same film were then exposed to $Ar^+$ ion irradiation with low and high fluence (2.6 and $7.0\times10^{14}$ ions/cm$^2$, respectively) and each with four different kinetic energies (1, 5, 10, 25 keV). Argon is chosen because it is nonmagnetic and non-reactive, well suited for intermixing in the upper layers due to its atomic weight. The energetic ions create local defects through atomic collisions, transferring energy to the host lattice. The irradiation depth is expected to increase with the ion energy and the amount of degradation scales with the fluence. The TRYDYN simulation software[22] was used to calculate the amount of layer intermixing and sputtering of the sample by the irradiation. Simulations show a depth profile for appreciable structural modification extending between 4-38 nm ($D_{\text{Irradiation}}$ in Table 1) for irradiation energies of 1-25 keV, deeper for higher energy irradiations, and a more thorough intermixing for the higher fluence.

Magnetic properties were measured by vibrating sample magnetometry (VSM), magneto-optical Kerr effect, and X-ray magnetic circular dichroism (XMCD), the latter using electron yield detection at the Advanced Light Source (ALS) on beamline 6.3.1. VSM and XMCD were performed following the first-order reversal curve (FORC) measurement protocol.[23-27] These measurements were correlated with structural characterizations by X-ray reflectivity (XRR) and X-ray diffraction (XRD) using a Bruker D8 Discover thin film diffractometer with Cu K$_\alpha$ radiation.

The FORC measurement protocol is useful for identifying details of the irreversible switching events. To measure a family of FORCs, the sample is first saturated. The field is then lowered to a reversal field ($H_R$) and the magnetization is measured back toward saturation as a function of the applied field ($H$). Executing this procedure for a series of reversal fields, a two dimensional map of the magnetization as a function of both the reversal field and the applied field, $M(H,H_R)$, is created and used to extract a FORC distribution:[28]

$$\rho(H, H_R) = -\frac{1}{2}\frac{\partial^2 M(H,H_R)}{\partial H \partial H_R}. \tag{1}$$

By integrating the FORC distribution along the applied field ($H$), we can obtain the FORC switching field distribution (FORC-SFD), which conveniently illustrates the extent of irreversible switching events as has been shown previously.[8, 29, 30]

XRR measurements show similar trends for both low and high fluence series, as shown in Figs. 1(a) and 1(b), respectively. The non-irradiated film and those irradiated with different kinetic energy $Ar^+$ all exhibit numerous high frequency oscillations below $2\theta \sim 4°$, which correspond to the total film thickness and indicate overall good layer structures. Except for the 10 and 25 keV high fluence samples, there are clear additional low frequency oscillations ($\sim 0.5°$ apart) in all other samples due to the Pd seed layer, as well as a superlattice peak near $2\theta \sim 7.8°$ due to the bilayer thickness of 1.2 nm. The decay of reflectivity oscillations as a function of angle is manifestation of an effective Debye-Weller factor with the layer roughness being a key parameter. The suppression of the XRR oscillations indicates an increase of interfacial roughness and a higher degree of disorders with increasing $Ar^+$ fluence and kinetic energy. Note that in the 1 keV samples the high frequency oscillations are suppressed at a lower angle compared to the 5 keV samples. This is because the entire ion energy is deposited into the top $\sim 5$ nm, leading to a more sever surface modification while leaving the layers underneath intact.

Magnetic hysteresis loops measured with the applied field perpendicular and parallel to the film plane exhibit clear perpendicular magnetic anisotropy, as shown for both the low [Figs.

2(a) and 2(c)] and high [Figs. 2(b) and 2(d)] fluence series. Out-of-plane loops of the films irradiated with 1 keV Ar$^+$ (black curve with solid squares) in both series show an 100% remanence and a precipitous magnetization drop upon nucleation and propagation of reversed domains, whereas the in-plane loops show a magnetic hard-axis behavior. As the irradiation energy increases, the perpendicular loop becomes more slanted, coercivity and remanent magnetization both decrease, and the nucleation field from positive saturation shifts to higher fields similar to effects reported in the literature for other graded media systems.[8, 12, 13] A consistent trend is also observed in the in-plane geometry [Figs. 2(c) and 2(d)], where the higher energy irradiated samples exhibit a larger in-plane remanent magnetization. For Ar$^+$ irradiation at the same kinetic energy, the higher fluence leads to a stronger reduction of the perpendicular anisotropy. These trends are due to the downward propagation of structural degradation under more energetic ion bombardments or a higher fluence. The destruction of the interface anisotropy causes the local magnetic easy axis to tilt towards the film plane.

FORC's have been measured by VSM and the corresponding FORC distributions calculated for all irradiated samples. For clarity, only those for a representative sample of the 10 keV high fluence film are shown in Fig. 3. The family of FORC's in Fig. 3(a) shows that irreversible switching starts to occur for $H_R < 0.2$ kOe, where the FORC's deviate from the major loop. This is more clearly seen in the three main features of the FORC distribution [Fig. 3(c)]: a horizontal ridge in -0.2 kOe $< H_R <$ 0.2 kOe, a largely reversible plateau in -1.8 kOe$< H_R <$ -0.2 kOe, and a positive vertical peak in -3.2 kOe $< H_R <$ -1.8 kOe, corresponding to the initial rapid nucleation and propagation of labyrinth domains, expansion/contraction of domains, and finally annihilation of residual domains to achieve saturation, as has been previously reported.[7, 24] The FORC measurement protocol has been expanded to XMCD measurements. Using the total electron yield signal by monitoring the sample drain current limits the magnetic sensitivity to the top 5-10 nm of the samples. This gives the unique ability to measure irreversible switching events within the *irradiated surface layers* of the Co/Pd multilayer, in contrast to the collective response from the entire film measured by VSM. For the high fluence 10 keV sample, the XMCD family of FORC's [Fig. 3(b)] is markedly different from the VSM counterpart [Fig. 3(a)], while the FORC distributions are still qualitatively similar [Fig. 3(d) vs. 3(c)]. This contrast is observed in several low fluence (10 and 25 keV) and high fluence (5 and 10 keV) samples, but absent in 1 keV (both fluencies) and low fluence 5 keV samples which exhibited essentially no differences in FORC's or FORC distributions between measurements of the surface (XMCD) and the whole film (VSM).

For samples irradiated with the high fluence 5 keV $Ar^+$ and all higher irradiation energies, a key difference in magnetization reversal occurs *prior to* the abrupt magnetization drop along the descending branch of the hysteresis loop (and the symmetric part along the ascending branch). The abrupt magnetization drop is caused by the nucleated domains rapidly propagating laterally across the thin film.[24] In the preceding region the successive FORC's lay on top of one another, as highlighted by dashed ovals in Figs. 3(a) and 3(b) for the high fluence 10 keV irradiated sample. The lowering of the magnetization is completely reversible, caused by canting of the moment towards the film plane. The much larger magnetization reduction in the surface (XMCD) compared to the bulk (VSM) measurement confirms that the canting starts from the irradiated surface, as a result of the weakened perpendicular anisotropy caused by the local structural damage. Once the nucleated domains start to propagate, the corresponding horizontal ridge in the FORC distribution occurs at nearly the same field values: the ridge in the XMCD-FORC distribution is centered at $H_R$ = 0.3 kOe [Fig. 3(d)] whereas that in the VSM-FORC distribution is centered at 0.1 kOe [Fig. 3(c)]. This comparison shows that the irreversible switching occurs in both the surface layer and the bulk of the film, with the surface layer leading the way. This is consistent with a vertical partial domain wall nucleating in the surface layer and propagating vertically through the film. At the onset of irreversible switching [indicated by arrows in Figs. 3(a) and 3(b)], the average canting angle of the surface ($\bar{\theta}_{surface}$) and bulk ($\bar{\theta}_{bulk}$) moments away from film normal are estimated using $\cos\bar{\theta} = M_N/M_S$, where $M_N$ is the perpendicular magnetization at the onset and $M_S$ is the saturation magnetization. As summarized in Table 1, all samples but the 1 keV and low fluence 5 keV samples (where the irradiation has not appreciably changed the magnetic anisotropy) show a higher surface canting angle than the bulk, and the angle increases with irradiation energy and fluence. This clearly demonstrates a vertical anisotropy grading that is more pronounced at a higher irradiation energy or fluence. For the high fluence 25 keV sample, the XMCD signal in the perpendicular geometry shows a hard axis character. This is evidence that the magnetic moments of the surface layer are lying completely in the plane of the film. It also indicates a higher density of pinning sites in the surface region due to defects.[7]

To illustrate the effect of irradiation on magnetization reversal, the VSM FORC-SFD is shown in Fig. 4, capturing the extent of the irreversible switching as the reversal field $H_R$ decreases. For the low fluence series [Fig. 4(a)], the onset of irreversible switching (when $dM/dH_R$ starts to become non-zero) shifts to higher fields at higher irradiation energies, while the end of the irreversible switching (when $dM/dH_R$ returns to zero) stays constant, near -3 kOe. This shows that domain nucleation and propagation occur easier for samples irradiated with higher energy ions as a result of the anisotropy reduction, but complete annihilation of residual domains requires virtually the same field for all samples since such residual domains are pinned by the highest anisotropy regions.[24] The high fluence series [Fig. 4(b)] shows an overall similar trend. An exception is observed in the 25 keV sample (open circles), where the onset of the irreversible switching is in the same position as the 10 keV sample, but the annihilation event is completed at a higher $H_R$ value.

In summary, we have demonstrated a convenient method to create graded anisotropy in Co/Pd multilayers using post-deposition $Ar^+$ ion irradiation. Tuning the energy and fluence of the $Ar^+$ ion irradiation is a viable method to tailor the depth dependent anisotropy gradient. For irradiations at the lowest energy and fluence, a strong perpendicular anisotropy remains intact. Conversely, samples irradiated at higher energies show a marked difference between surface and bulk magnetization, confirming a vertical anisotropy gradient. FORC distributions show that the irradiated surface layer remains coupled to the bulk of the film, while facilitating an early nucleation of the reversed domains. An average canting angle of the magnetic moments in the surface layer is extracted, illustrating a magnetization reversal process via vertical partial domain wall propagation. These results show a promising alternative approach, which may be combined with the more traditional synthesis-based methods, to maximize the anisotropy gradient and fully realize the benefits of graded media.

This work has been supported by the U.S. National Science Foundation (DMR-1007891 & ECCS-0925626) at UCD and by the Deutsche Forschungsgemeinschaft (Grant No. FA 314/3-2) at HZDR. The work at the ALS was supported by the Director, Office of Science, Office of Basic Energy Sciences of the U.S. Department of Energy (DEAC02-05CH11231).

**Figure Captions**

**Figure 1**: X-ray reflectivity of (a) low and (b) high fluence irradiated films, along with the non-irradiated film as reference. The spectra are vertically offset for clarity.

**Figure 2:** VSM hysteresis loops of $Ar^+$ ion irradiated Co/Pd films, with field applied (a,b) perpendicular or (c,d) parallel to the film plane for (a,c) low and (b,d) high fluences.

**Figure 3:** (a,b) Families of FORC's and (c,d) the corresponding FORC distributions measured by (a,c) VSM and (b,d) XMCD for the high fluence 10 keV irradiated film with field applied perpedicular to the film plane. Dashed ovals in (a,b) highlight the initial reversible magnetization decline, where the onset of irreversible switching is illustrated by the arrows.

**Figure 4:** FORC switching field distributions for the (a) low and (b) high fluence series.

**Table. 1.** Average canting angle at the onset of irreversible switching estimated from normalized magnetization values, measured by VSM and XMCD, along with the irradiation depth $D_{\text{Irradiation}}$ calculated from TRYDYN.§

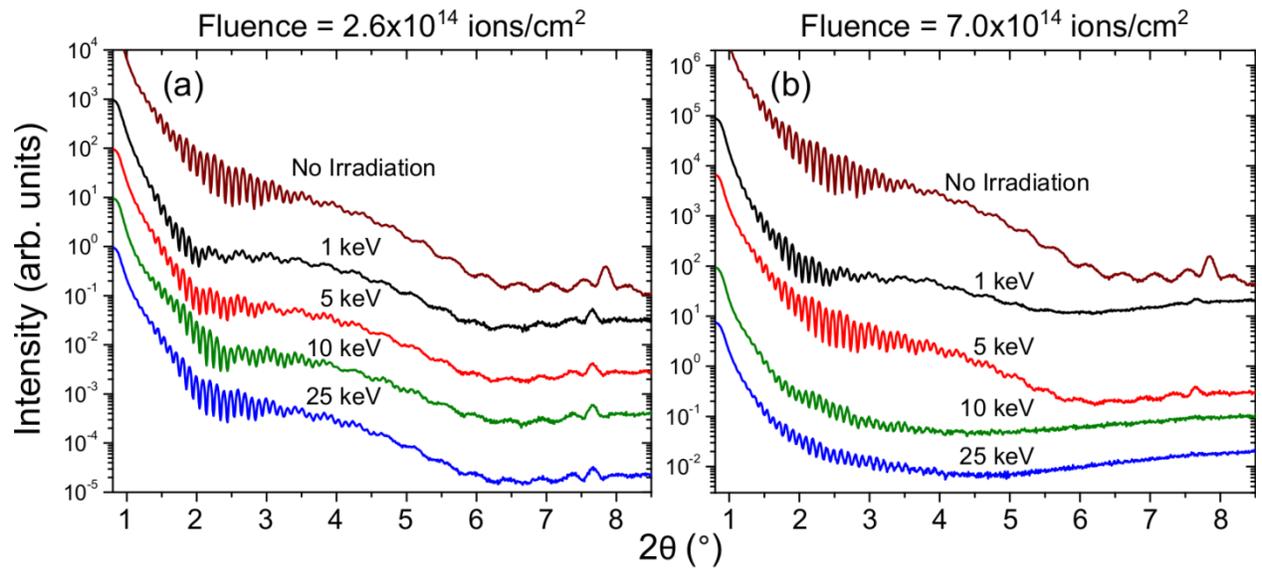

Fig. 1. Greene, *et al.*

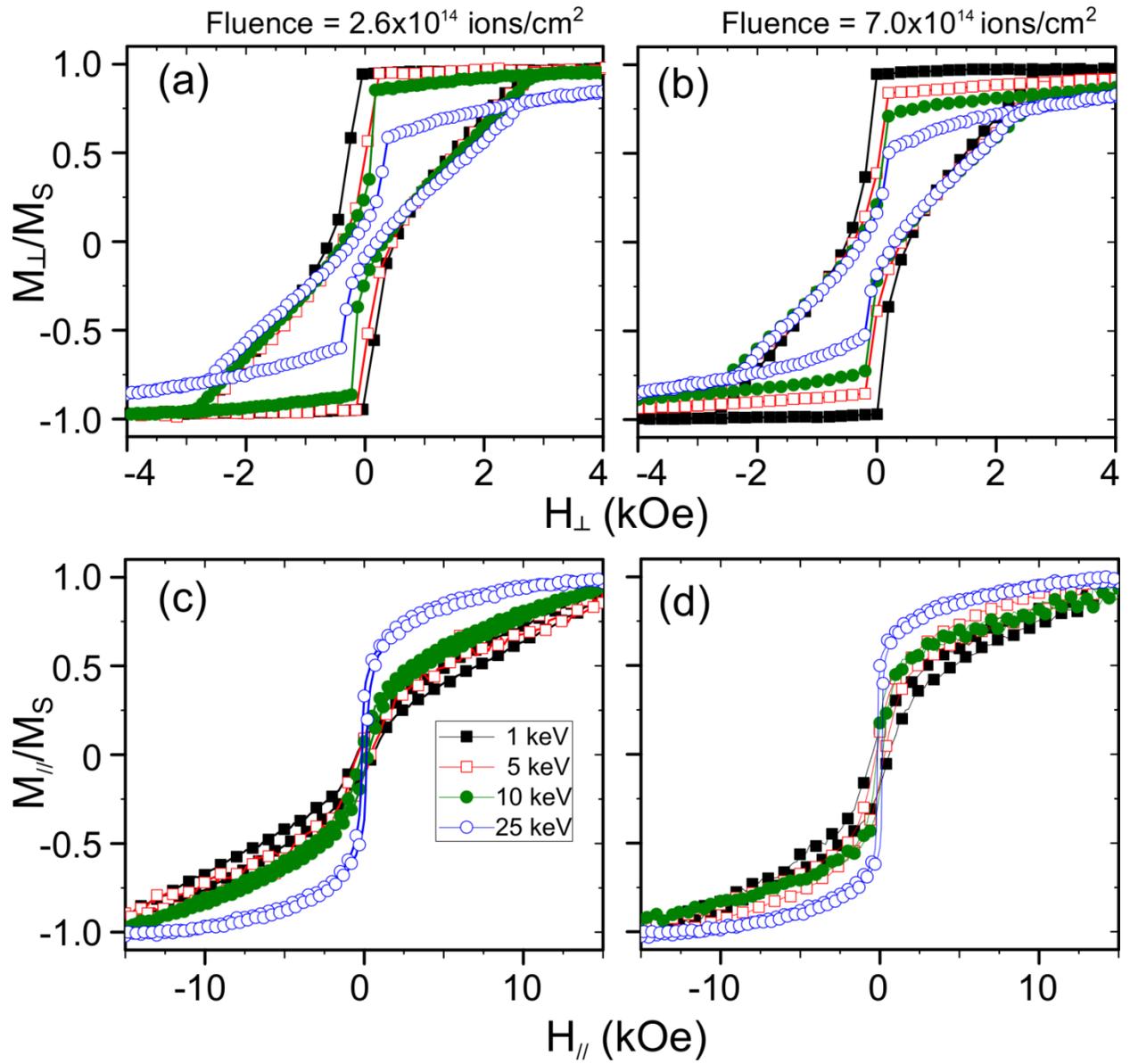

**FIG. 2. Greene,** *et al.*

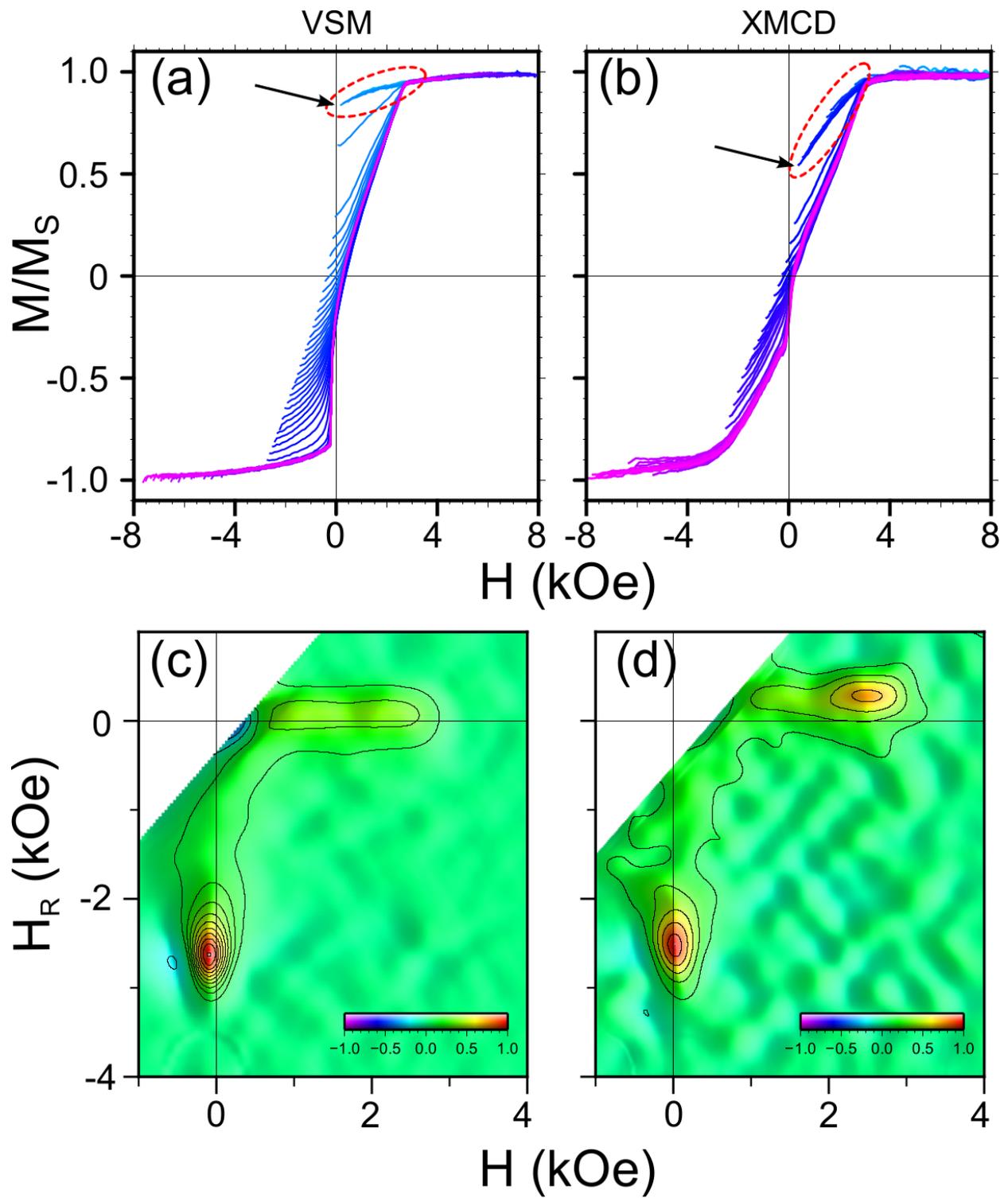

**FIG. 3.** Greene *et al.*

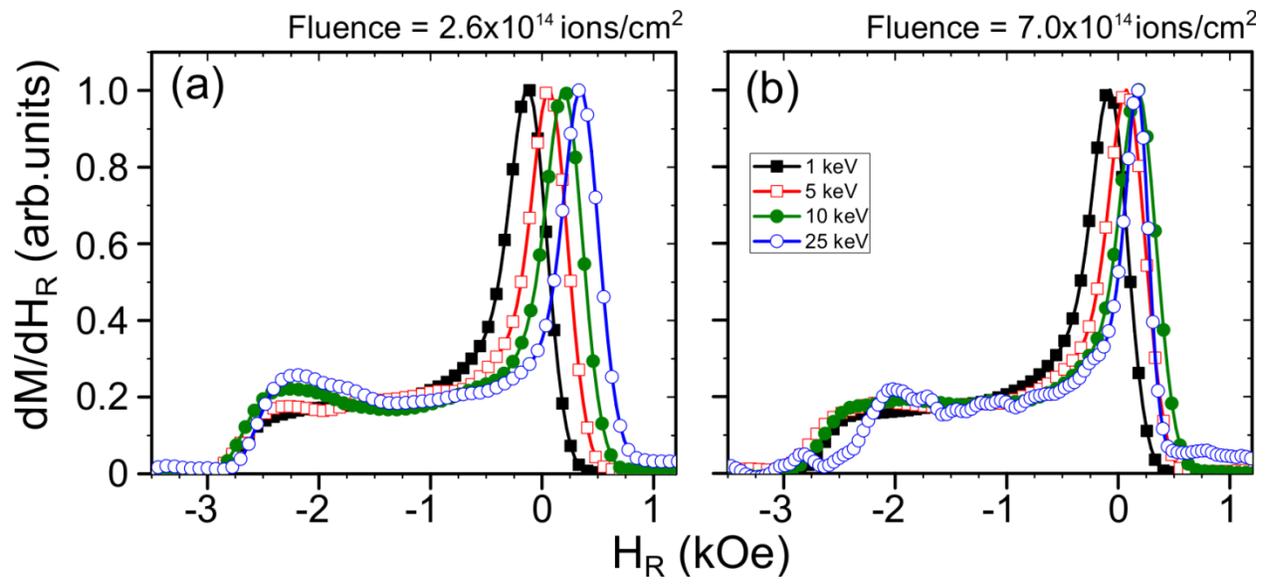

**FIG. 4.** Greene et al.

**Table. 1.** Average canting angle at the onset of irreversible switching estimated from normalized magnetization values, measured by VSM and XMCD, along with the irradiation depth $D_{\text{Irradiation}}$ calculated from TRYDYN.[§]

| $E$ (keV) | $D_{\text{Irradiation}}$ (nm) | Fluence ($10^{14}/\text{cm}^2$) | VSM $M_N/M_S$ Bulk | $\bar{\theta}_{Bulk}$ (°) | XMCD $M_N/M_S$ Surface | $\bar{\theta}_{Surface}$ (°) |
|---|---|---|---|---|---|---|
| 1 | 4.6±0.2 | 2.6 | 0.93 | 21 | 1.0 | 0 |
| 5 | 10.6±0.2 | 2.6 | 0.94 | 20 | 0.94 | 20 |
| 10 | 16.6±0.2 | 2.6 | 0.85 | 32 | 0.55 | 57 |
| 25 | 33.4±0.2 | 2.6 | 0.58 | 55 | 0.33 | 71 |
| 1 | 4.7±0.2 | 7.0 | 0.94 | 20 | 1.0 | 0 |
| 5 | 12.9±0.2 | 7.0 | 0.84 | 33 | 0.57 | 55 |
| 10 | 18.9±0.2 | 7.0 | 0.71 | 45 | 0.32 | 71 |
| 25 | 38.2±0.2 | 7.0 | 0.50 | 60 | N/A | N/A |

[§]Determined from intermixing of Co into Pd-layer with a concentration above 0.01%.